\documentclass[prl,showpacs,twocolumn,amsmath]{revtex4}
\include{Holedefs}  
\graphicspath{{./figures/}}
\begin{document}

\title{Coherent Control and Suppressed Nuclear Feedback of a Single Quantum Dot Hole Qubit}

\newcommand\ginzton{E. L. Ginzton Laboratory,
        Stanford University,
        Stanford, California 94305, USA}
\newcommand\NII{National Institute of Informatics,
        Hitotsubashi 2-1-2, Chiyoda-ku,
        Tokyo 101-8403, Japan}
\newcommand\wurzburg{Technische Physik, Physikalisches Institut,
        Wilhelm Conrad R\"{o}ntgen Research Center for Complex Material Systems,
        Universit\"{a}t W\"{u}rzburg,
        Am Hubland, D-97074 W\"{u}rzburg, Germany}
\newcommand\HRL{ Currently at HRL Laboratories, LLC, 3011 Malibu
        Canyon Rd., Malibu, CA 90265, USA}

\author{Kristiaan~De~Greve}
\affiliation\ginzton
\author{Peter~L.~McMahon}
\affiliation\ginzton
\author{David~Press}
\affiliation\ginzton
\author{Thaddeus~D.~Ladd}
\altaffiliation\HRL
\affiliation{\ginzton}\affiliation{\NII}
\author{Dirk~Bisping}
\affiliation\wurzburg
\author{Christian~Schneider}
\affiliation\wurzburg
\author{Martin~Kamp}
\affiliation\wurzburg
\author{Lukas Worschech}
\affiliation\wurzburg
\author{Sven~H\"{o}fling}
\affiliation\ginzton
\affiliation\wurzburg
\author{Alfred~Forchel}
\affiliation\wurzburg
\author{Yoshihisa~Yamamoto}
\affiliation{\ginzton}\affiliation{\NII}

\begin{abstract}
Future communication and computation technologies that exploit quantum information require robust and well-isolated qubits. Electron spins in III-V semiconductor quantum dots, while promising candidates, see their dynamics limited by undesirable hysteresis and decohering effects of the nuclear spin bath. Replacing electrons with holes should suppress the hyperfine interaction and consequently eliminate strong nuclear effects. Using picosecond optical pulses, we demonstrate coherent control of a single hole qubit and examine both free-induction and spin-echo decay. In moving from electrons to holes, we observe significantly reduced hyperfine interactions, evidenced by the reemergence of hysteresis-free dynamics, while obtaining similar coherence times, limited by non-nuclear mechanisms. These results demonstrate the potential of optically controlled, quantum dot hole qubits.\\
\end{abstract}

\maketitle

The coherence of quantum bits (qubits) is crucial for the implementation of quantum computation and communication schemes~\cite{thaddeusQC}. Electron spins in III-V semiconductor quantum dots, among the fastest, most promising qubits, see their coherent dynamics limited by strong hyperfine interactions with the nuclear spin bath~\cite{loss04,petta05,sham06,witzeldecoherence,daveT2}. Dynamical decoupling, the repeated application of rotation pulses to decouple a qubit from a noise bath, has been applied to the electron-nuclear hyperfine interaction, and was shown to increase the coherence of a single electron spin~\cite{bluhmdd}. Unfortunately, electron-nuclear spin bath dynamics are highly non-Markovian~\cite{vinknuclei,lattanuclei,steel09,thaddeusnuclei}, a cumbersome feature which complicates the control
operations required for effective dynamical decoupling, and which affects the fidelity of control operations~\cite{vinknuclei}. Isotopic engineering may also reduce nuclear noise. While this is possible for quantum dots or charged impurities in group-IV or group II-VI semiconductors~\cite{thaddeusQC}, III-V semiconductor compounds do not possess stable spin-zero isotopes. The strong electron-spin contact hyperfine interaction, however, can be avoided by engineering qubits based on valence-band holes~\cite{Bulaev,fischerprb,gerardot08,holecpt}, as the symmetry of the hole wavefunction leaves only the weak residual dipolar interaction as the leading term in the hyperfine Hamiltonian~\cite{fischerprb}.

Electrons and holes are not completely equivalent, but as we show here, they can be controlled by the same methods. The large heavy-hole (HH)-light-hole (LH) splitting in III-V QDs makes it possible to consider only the HH manifolds, especially at low temperatures. The HHs can be described in a pseudo-spin-$1/2$ formalism~\cite{Bulaev}, with a perturbative treatment of LH inmixing. The \textit{p}-like symmetry of the hole Bloch wavefunction enables more spin-orbit coupling mechanisms than for an electron spin~\cite{Bulaev}, especially in the case of significant HH-LH mixing~\cite{chargenoise}. The hole-spin decoherence is therefore more sensitive to electric fields and orbital degrees of freedom than that of electron spins. We nevertheless confirm in this work that the physical mechanisms underpinning optical initialization and ultrafast optical control of hole qubits are essentially identical to those of electron spins~\cite{bmsca08,press08}, and allow for an arbitrary single-qubit rotation to be performed in several tens of picoseconds. This is in contrast to RF and microwave control of hole qubits: there, the coupling of a single HH with the control field is significantly lower than that of an electron spin~\cite{darinholes}, resulting in orders of magnitude slower control operations. Alternatives such as the use of LH spins (through strain engineering of III-V heterostructures), exchange coupling between HH spins in a double QD, or all-electrical QD-molecule HH spin $g$-tensor control have been proposed~\cite{darinholes,roloff}, though have yet to be implemented.

\begin{figure}\begin{center}
\includegraphics[width=0.9\columnwidth]{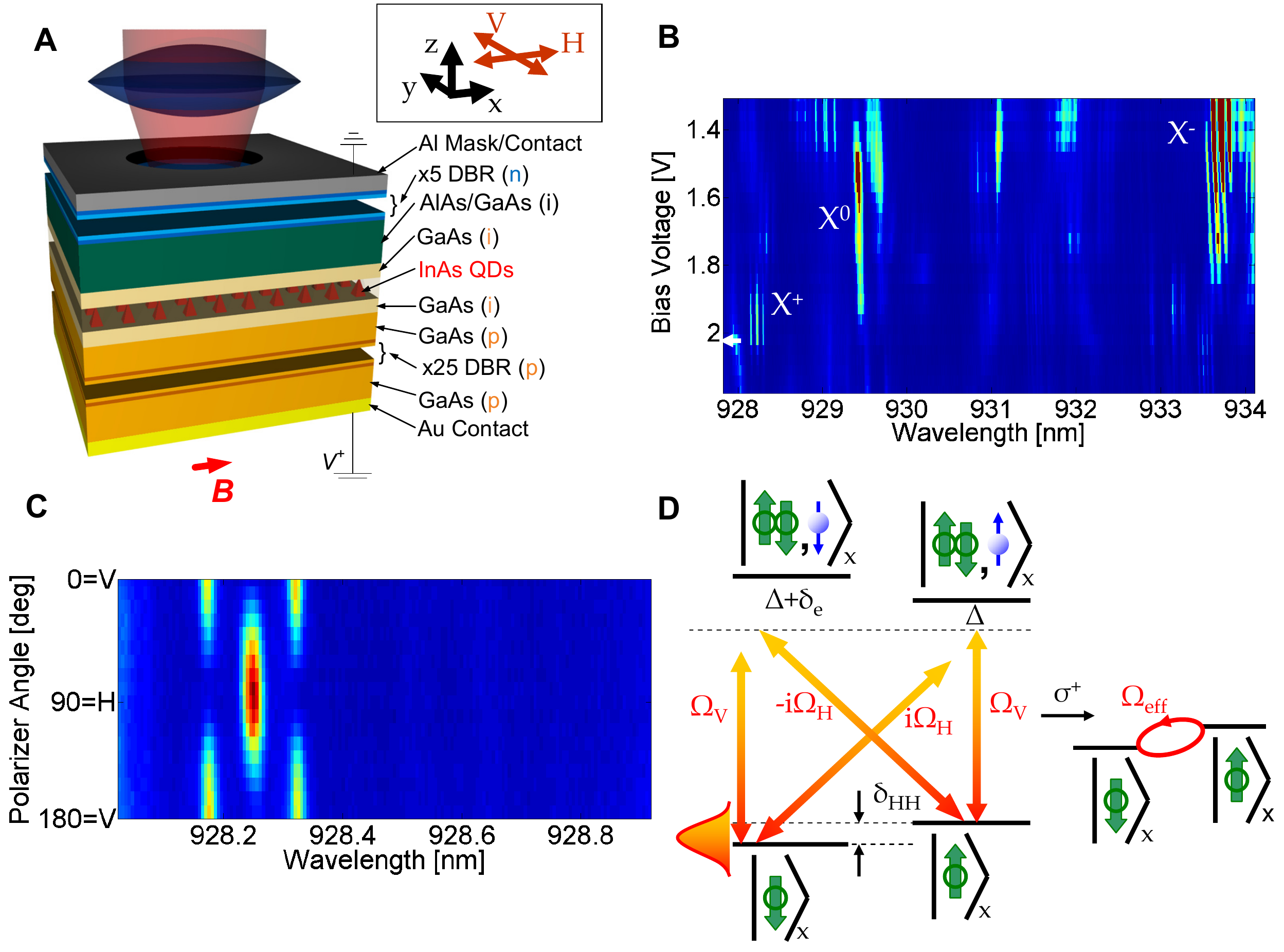}
\end{center}
\vspace{-25pt}
\caption{
(\textbf{A})~Sample structure and geometry used (not to scale). A single layer of quantum dots is embedded in a p-i-n-diode, and separated from a hole reservoir by a 25 nm tunnel barrier. A 110 nm AlAs/GaAs superlattice prevents tunneling from an electron reservoir. Distributed Bragg Reflector pairs (DBR, 5 pairs on top and 25 pairs on the bottom) form a low-\textit{Q} cavity. We isolate a single QD through spatial and spectral filtering. A magnetic field $B$ is applied along the $x$-axis (Voigt geometry), and a bias voltage $V^{+}$ determines the charge state of the QD. Inset: orientation (x,z) and polarization convention (H,V) used.
(\textbf{B})~Magnetophotoluminescence plot ($B=6$~T; above-band pumping, $\lambda=785$~nm) of a QD used in the experiment, as a function of the bias voltage. X$^{+}$ indicates the hole-charged trion state, X$^{-}$ the optical-excitation induced, electron-charged trion, and X$^{0}$ the neutral exciton state. The white arrow indicates the bias voltage used in Fig.~1\textbf{c}.
(\textbf{C})~Polarization dependence of the X$^{+}$ photoluminescence when biased at 2.0 V ($B=6$~T). Note that two of the four X$^{+}$-emission lines overlap at this magnetic-field strength, leading to the apparent increase in brightness of the center lines.
(\textbf{D})~Level structure with idealized polarization selection rules, as used in the experiment. Applying a detuned broadband laser pulse ($\Delta$/2$\pi=340$~GHz, 3.67 ps FWHM) generates an effective coupling between the HH-states ($\Omega_{\mathrm{eff}}$). For an 8~T magnetic field, $\delta_{\mathrm{HH}}$/2$\pi=30.2$~GHz, $\delta_{\mathrm{e}}$/2$\pi\sim35$~GHz.
\vspace{1pc}
}
\end{figure}

We systematically studied coherent single-hole manipulation in several QDs in multiple devices, using two different types of device structures (charge-tuneable and hole-$\delta$-doped~\cite{SOM}). The results from different QDs and devices are qualitatively equivalent, although quantitative measurements do yield different values from dot to dot.  All figures in this Report result from one particular charge-tuneable QD, which we deterministically charge with a hole by embedding it in a p-i-n-diode (Fig.~1\textbf{A}) - an approach similar to the one used in Refs.~\onlinecite{warburtoncharging},\onlinecite{gerardot08},\onlinecite{holecpt}. We bias the diode so as to load a single hole into the QD, which we detect optically through magneto-photoluminescence (Fig.~1\textbf{B}). A magnetic field is applied in Voigt geometry (perpendicular to the optical axis, Fig.~1\textbf{A}), giving rise to a double $\Lambda$-system under optical excitation (Figs.~1\textbf{C} and \textbf{D}). Previous work on electron spins~\cite{bmsca08,press08} demonstrated the coherent and ultrafast manipulation of such a $\Lambda$-system with circularly polarized, detuned, picosecond pulses. We employ a similar ultrafast coherent manipulation scheme for the HH in our QD (Fig.~2\textbf{A}). A magnetic field of 8~T splits the HH eigenstates $\left|\Downarrow \right\rangle$ and $\left|\Uparrow \right\rangle$ by the pseudo-spin Larmor frequency $\delta_{\mathrm{HH}}$/2$\pi=30.2$~GHz. Before applying any rotation pulses, we initialize the hole qubit by optical pumping~\cite{atature06,Xuprl,daveT2,gerardot08}, for which we use the hole-trion state ($\left|\Downarrow \Uparrow, \uparrow \right\rangle$, consisting of a HH singlet and an unpaired electron spin). A 26~ns narrowband continuous wave (CW)-laser pulse (1-2~MHz linewidth before modulation) is applied resonant with the $\left|\Uparrow \right\rangle$-$\left|\Downarrow \Uparrow, \downarrow \right\rangle$-transition, which initializes the hole spin into the $\left|\Downarrow \right\rangle$-state in a few ns. The same CW-laser pulse sequence is used for reading out the hole spin state. If, after initialization, the coherent manipulation pulse rotates the spin into the  $\left|\Uparrow \right\rangle$-state, then the subsequent optical-pumping pulse will cause a single photon to be emitted from the $\left|\Downarrow \Uparrow, \uparrow \right\rangle$-$\left|\Downarrow \right\rangle$-transition. This photon is filtered, and subsequently detected by a single-photon counter. For coherent manipulation, we use broadband pulses (FWHM:  3.67~ps) from a modelocked laser, detuned by $\Delta$/2$\pi=340$ GHz. A combination of electro-optic modulators, beamsplitters and variable-delay paths (Fig.~2\textbf{B}) allows accurate control over the timing of the pulses~\cite{SOM}.

\begin{figure}\begin{center}
\includegraphics[width=0.9\columnwidth]{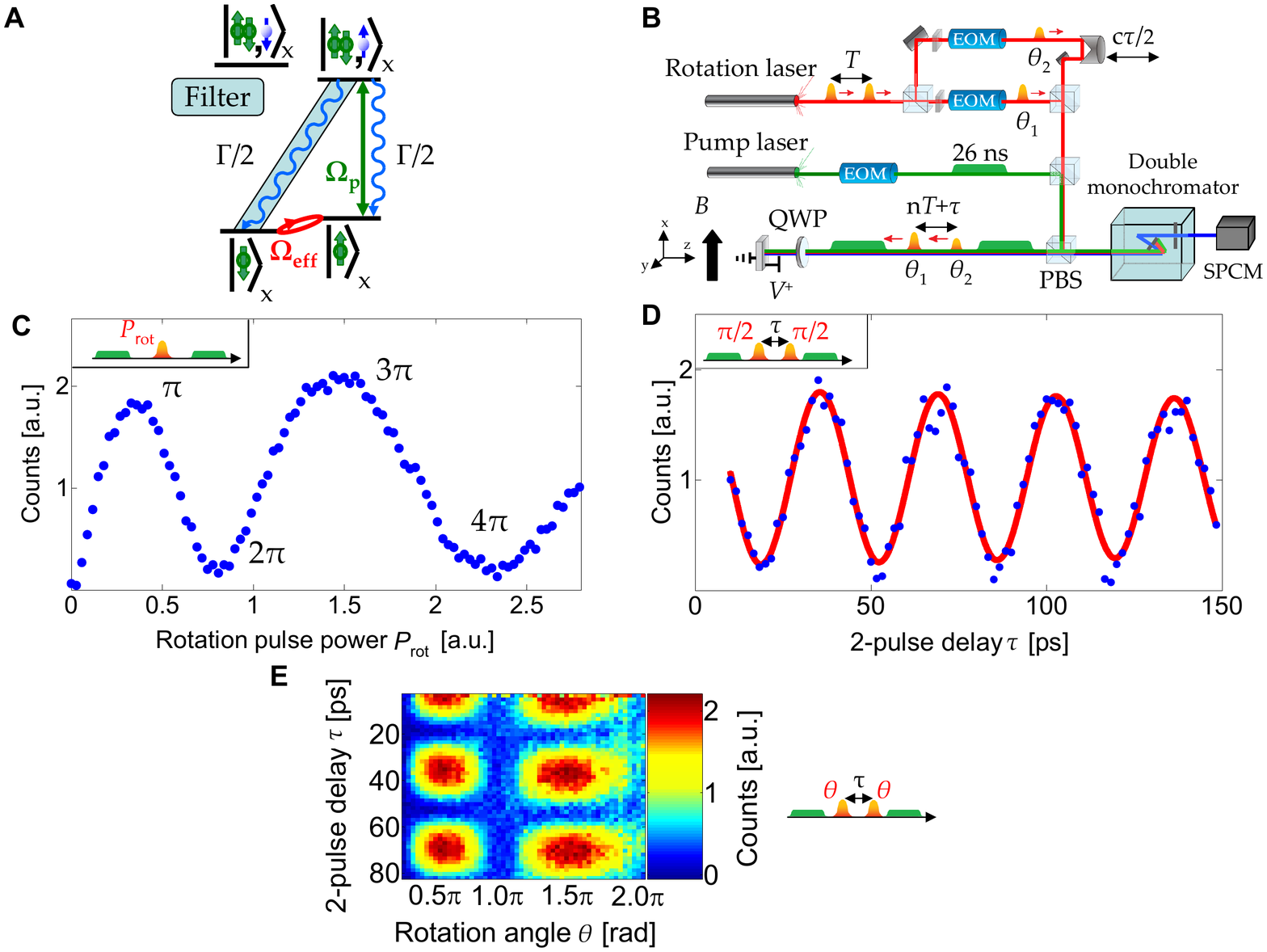}
\end{center}
\vspace{-25pt}
\caption{
(\textbf{A})~Schematic overview of the all-optical manipulation scheme. $\Gamma$: spontaneous emission rate from the $\left|\Downarrow \Uparrow, \downarrow \right\rangle$-state, estimated at (1~ns)$^{-1}$; $\Omega_{\mathrm{p}}$:  coupling due to the optical pumping/readout laser; $\Omega_{\mathrm{eff}}$: effective Rabi-coupling resulting from the detuned modelocked-laser pulses.
(\textbf{B})~Experimental setup. The ML-laser pulses (period: $T=13$~ns) are split into two independently adjustable (power, delay $\tau$) branches, allowing for multiple rotation laser pulses to be applied sequentially. A double monochromator is used to filter out the scattered laser light.
    QWP: quarterwave plate;
    PBS: polarizing  beamsplitter;
    EOM: electro-optic modulator;
    SPCM: single photon counting module;
    $V^{+}$: bias voltage;
    $B$: external magnetic field.
(\textbf{C})~Coherent Rabi oscillations. When varying the power $P_{\mathrm{rot}}$ of the rotation pulses, the HH-spin state coherently evolves from $\left|\Downarrow\right\rangle$ to $\left|\Uparrow \right\rangle$ and back, resulting in an oscillatory photon count signal. Inset: the timing of the initialization/readout pulses (green) and rotation pulse (red).
(\textbf{D})~Ramsey interference fringes. After a first $\pi$/2 rotation pulse, the hole spin is allowed to precess around the external magnetic field (Larmor precession, $\omega_{\mathrm{L}}$/2$\pi = 30.2$~GHz). By delaying a second $\pi$/2 pulse by an amount $\tau$, Ramsey fringes are observed. Inset: timing of the initialization/readout and rotation pulses.
(\textbf{E})~Demonstration of complete SU(2) control of the single hole qubit. By changing both the delay $\tau$ and the pulse rotation angle $\theta$, the entire surface of the Bloch sphere is explored. Inset: timing and pulse amplitude used.
}
\end{figure}

The effect of a single rotation pulse is outlined in Fig.~2\textbf{C}. The rotation pulse effectively couples the $\left|\Downarrow \right\rangle$ and $\left|\Uparrow \right\rangle$-states, which can be equivalently interpreted as implementing a stimulated Raman transition~\cite{esws06,press08} or as the result of an AC-Stark-shift~\cite{bmsca08}. For a fixed rotation pulse duration, Rabi oscillations between $\left|\Downarrow \right\rangle$ and $\left|\Uparrow \right\rangle$ are observed as a function of the rotation pulse power.

The Rabi oscillations demonstrate single qubit rotations around the $z$-axis (Fig.~1\textbf{A} shows the axis conventions used). However, arbitrary qubit control requires controlled rotation about a second axis. For this, we use the effective Larmor precession of the HH pseudo-spin around the magnetic-field axis ($x$-axis). This precession can be probed through Ramsey interferometry. In Fig.~2\textbf{D}, the resulting Ramsey fringes are shown when two $\pi$/2 pulses are applied with a variable delay $\tau$. From the Ramsey fringe visibility, a fidelity of $0.945$ can be deduced for a single $\pi/2$ rotation. Any arbitrary single qubit rotation can be performed by decomposition into rotations about the $x$- and $z$-axes~\cite{press08}. This is illustrated in Fig.~2\textbf{E}, where the entire Bloch sphere surface is explored by sweeping through both the $z$-rotation angle ($\theta$) and $x$-rotation angle ($\delta_{\text{HH}}\tau$). Using these methods, any single qubit rotation can be performed in  approximately 20~ps or less~\cite{press08}.

\begin{figure}\begin{center}
\includegraphics[width=0.9\columnwidth]{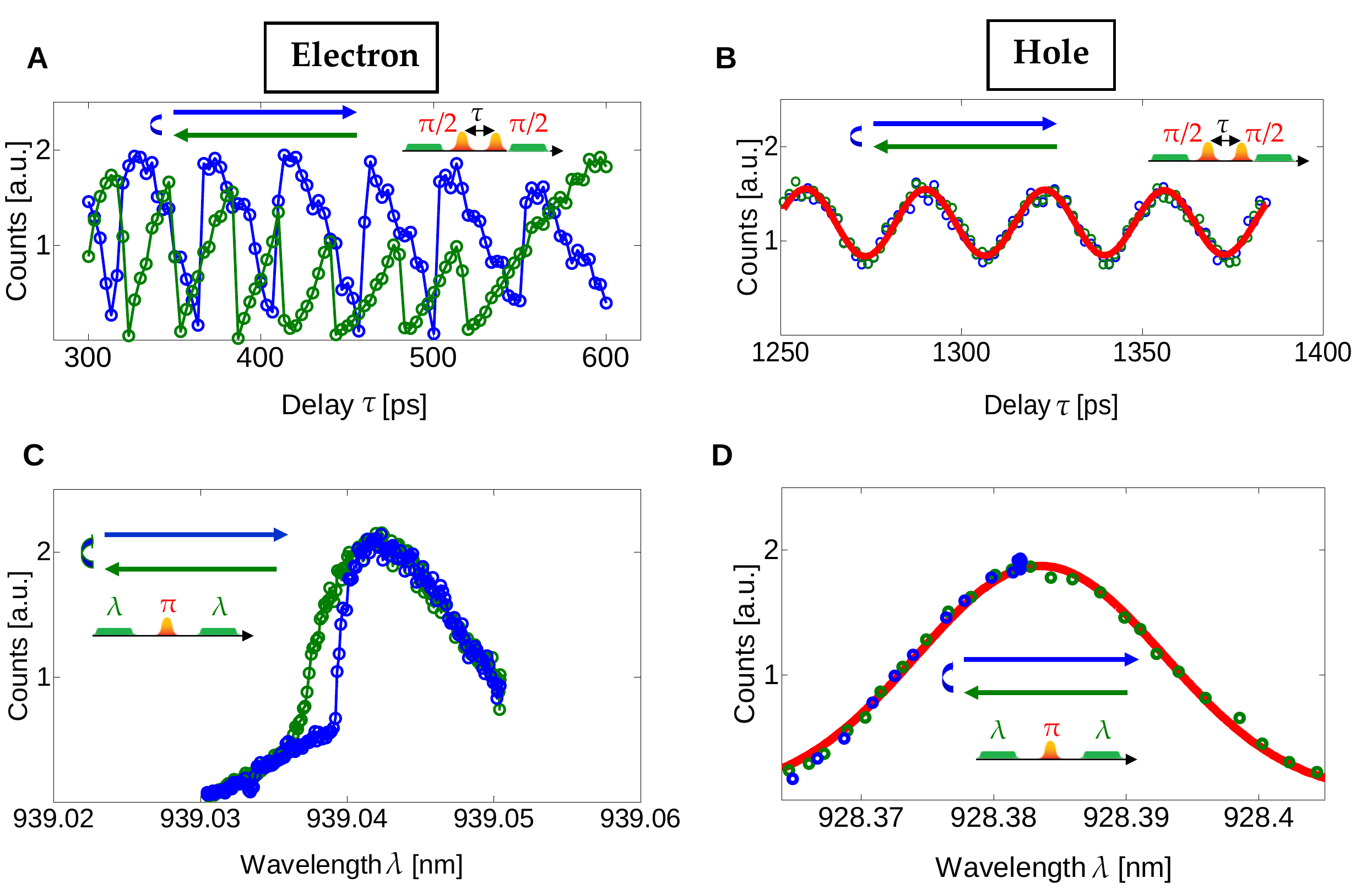}
\end{center}
\caption{
(\textbf{A})~Asymmetric and hysteretic Ramsey fringes for an electron spin in a QD. The green and blue circles refer respectively to backwards and forwards scanning of the delay line, as indicated by the arrows~\cite{thaddeusnuclei}. Inset: the pulse sequence used in the experiment.
(\textbf{B})~Symmetric and hysteresis-free Ramsey fringes for a hole spin in a QD. Even for large delays, the Ramsey fringes remain perfectly sinusoidal. The green and blue circles refer to different scanning directions, as indicated by the arrows. The red curve represents a sinusoidal least-squares fit. Inset: pulse sequence used.
(\textbf{C})~Effective absorption signal for an electron spin in a QD, showing clear asymmetry and hysteresis upon scanning the pump laser wavelength in different directions (green and blue circles, as indicated by the arrows). Inset: pulse sequence used.
(\textbf{D})~Effective absorption signal for a hole spin in a QD. No hysteresis or asymmetry was observed upon scanning in different directions (green and blue circles, as indicated by the arrows). Red curve: least-squares fit of a Gaussian absorption profile. Inset: pulse sequence used.
}
\end{figure}

The complete SU(2) control of a QD spin allows for the observation of hyperfine interactions between the spin and the nuclei. Nuclear spin interactions manifest themselves primarily as Overhauser shifts of the effective magnetic field.  These cause both dephasing of qubit memory and angle errors in single qubit control. In principle, compensation methods exist for static Overhauser shifts, but such shifts change over time, and unfortunately their dynamics have been shown to depend on previously applied control operations.  For electron spins in single QDs, such non-Markovian, hysteretic behavior was observed in both Ramsey interferometry~\cite{thaddeusnuclei} and in continuous-wave (CW) laser scans~\cite{steel09,lattanuclei}. A suppressed hyperfine interaction for a hole qubit should show an absence of these effects, and allow for easy implementation of accurate $x$-rotations in arbitrary pulse sequences.

  In Figs.~3\textbf{A} and \textbf{B}, we compare the analogs of free-induction decay (FID) for a single electron spin and a hole pseudo-spin through Ramsey interferometry~\cite{thaddeusnuclei}. For a single electron spin, the resulting Ramsey fringes are shown in Fig.~3\textbf{A}, using experimental parameters similar to those reported previously~\cite{thaddeusnuclei}. For these experiments, the nuclear spin polarization is significantly altered during each optical pumping cycle. Since the nuclear polarization affects the electron spin's coherent evolution, a strong feedback loop develops, resulting in the observed hysteretic behavior~\cite{thaddeusnuclei,steel09}. For a HH qubit, shown in Fig.~3\textbf{B}, no such hysteresis was observed. This lack of observable hysteresis is understood to be due to the reduction of Overhauser shifts that results from the absence of a contact hyperfine coupling for holes. We observe a reduction of least 30 in the nuclear feedback strength when using holes instead of electrons, in line with recent direct measurements of the reduced hyperfine coupling of holes~\cite{imamogluholes,tartaholes}. This estimate is a conservative lower bound, limited by hole decoherence that may obscure possible weak effects~\cite{SOM}.

We also performed experiments in which we scan the optical pumping laser while applying a $\pi$-rotation pulse (inset of Figs.~3\textbf{C} and \textbf{D}). This mimics two-CW-laser resonant absorption experiments performed for electrons~\cite{lattanuclei,steel09} and for holes~\cite{holecpt,imamogluholes}. For electron spins, multiple experiments~\cite{thaddeusnuclei,lattanuclei,steel09} report that scanning the laser through the $\left|\downarrow \right\rangle$-$\left|\downarrow \uparrow, \Uparrow \right\rangle$-resonance leads to a hysteretic nuclei-induced wandering of that resonance; this effect is seen in Fig.~3\textbf{c}. In contrast, no hysteresis is observed for a hole-charged QD (Fig.~3\textbf{D}). The absorption profile is completely symmetric, and is best fit by a Gaussian (red curve in Fig.~3\textbf{D}) with a linewidth of 6.7~GHz. The notable broadening suggests significant spectral diffusion of our hole-charged QD. For different dots, as well as for measurements on the $\delta$-doped sample, we obtain comparable linewidths, typically 3 times larger than for similar, electron-doped QDs when measured in the same setup. Similar broadening of hole-charged QDs has been observed previously~\cite{holecpt}.



\begin{figure}\begin{center}
\includegraphics[width=0.9\columnwidth]{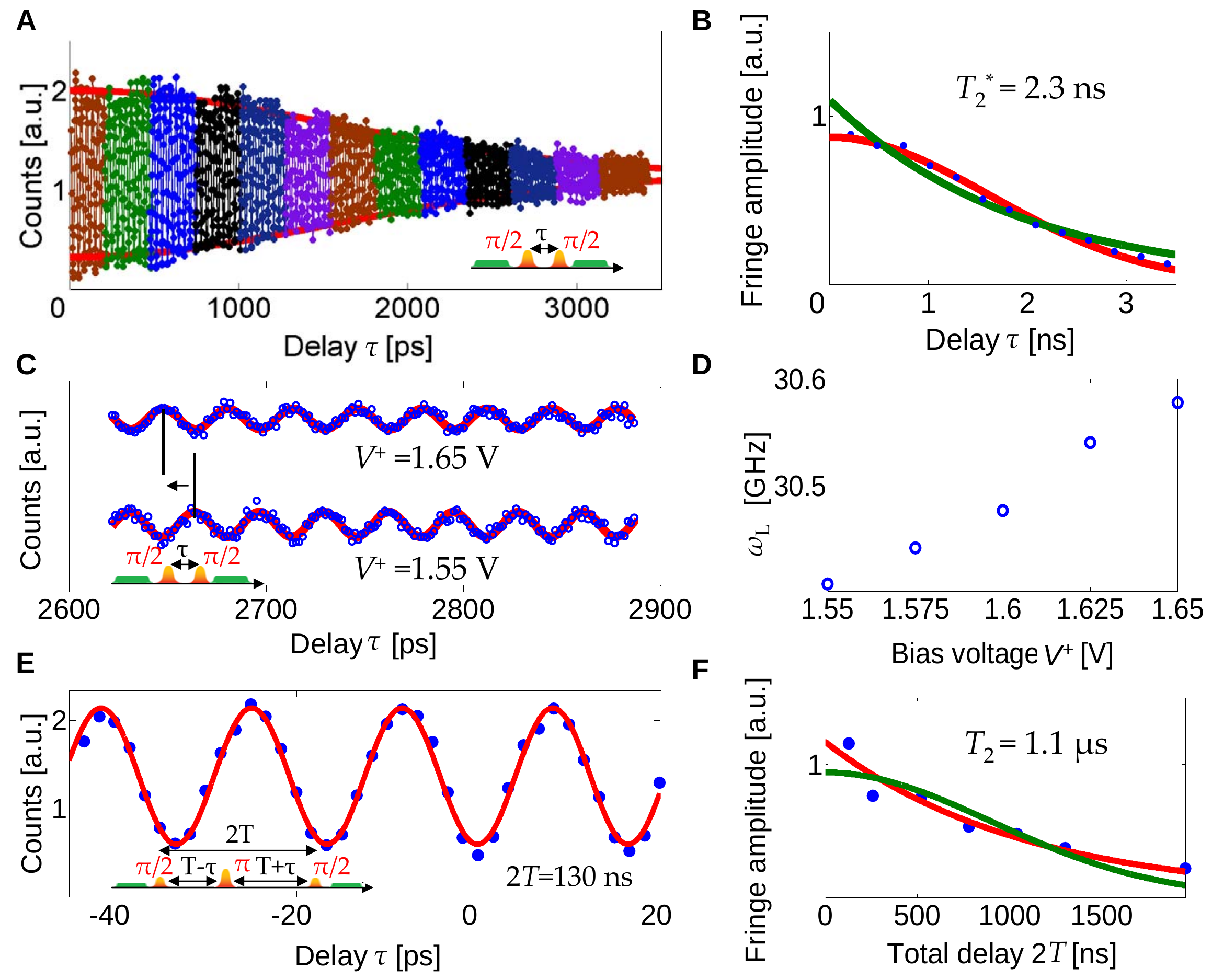}
\end{center}
\caption{
(\textbf{A})~Ramsey fringes as a function of the delay $\tau$ between two $\pi$/2 pulses. The different colors refer to different positions of the stage on the rail (see text), while the red envelopes indicate a least-squares-fit Gaussian decay with $T_{2}^{*}=2.3$~ns. Inset: pulse timing used.
(\textbf{B})~Amplitude of the Ramsey fringes, as a function of the delay $\tau$. Red curve: fit to a Gaussian decay ($T_{2}^{*}=2.3$~ns); green curve: fit to an exponential decay.
(\textbf{C})~Long-delay Ramsey fringes, for different applied voltage bias $V^+$. Top: $V^+$=1.65~V; bottom: $V^+$=1.55 V. Red curves: least-squares, sinusoidal fit. The cumulative effect of the difference in Larmor precession frequency $\omega{}_{\mathrm{L}}$ can be seen in the shifting of the curves.
(\textbf{D})~Larmor-precession frequency $\omega_{\mathrm{L}}$ as a function of the externally applied bias voltage $V^+$.
(\textbf{E})~Spin-echo fringes as a function of fine-delay $\tau$, for a total delay 2$T=130$~ns. Red curve: fit to a sinusoid. Inset: pulse timing used.
(\textbf{F})~Spin-echo fringe amplitude as a function of the total delay 2$T$. Red curve: fit to exponential decay ($T_{2}$=1.1$\mu$s); green curve: fit to Gaussian decay.
}
\end{figure}


A qubit's coherence is characterized by both $T_2^*$, which results from very low frequency noise, and by $T_2$, which characterizes decoherence due to higher frequency noise~\cite{thaddeusQC}. $T_2^*$ is found as the decay of the time-averaged hole-pseudo-spin FID, as shown in Fig.~4\textbf{A}. In Fig.~4\textbf{B} the fringe height is shown as a function of the delay between the pulses. The decay is best fit by a Gaussian (i.e., $\propto\exp[-(t/T_2^*)^2]$) with $T_{2}^* = 2.3$~ns, and is independent of the magnetic field for fields between 6 and 10~T. It is unlikely that nuclear spin effects are responsible for the $T_{2}^{*}$ times we observe. Theoretical calculations of the FID of a single hole predict much longer time-averaged dephasing times when limited by hyperfine interactions, even when taking finite HH-LH mixing into account~\cite{fischerprb,Fischermixed}. Non-nuclear dephasing processes are far more likely; such processes limit coherence in localized hole spins in quantum wells~\cite{chargenoise,chargenoiseBayer}, and may even overwhelm nuclear processes in electron-charged quantum dots at some magnetic fields, depending on the dot's proximity to noisy surface states~\cite{daveT2}. These dephasing processes generally arise from fluctuations in the localizing potential, a process that may be observed as spectral diffusion of the optical transitions, evidenced by the optical linewidth in the resonance-scanning experiment described above.

The effect of spectral diffusion on hole spin coherence, in the form of randomly varying electric fields, can be directly examined in our sample. By deliberately changing the electrical bias over the QD, corresponding to spectral shifts similar in magnitude to those presumedly responsible for spectral diffusion, we measure notably
different effective Larmor precession frequencies of the HH pseudo-spin. In Fig.~4\textbf{C}, the cumulative effect of such a difference in Larmor frequency is shown through long-delay Ramsey fringes. The different precession frequencies lead to anti-phase Ramsey fringes for delays similar to $T_{2}^{*}$. In Fig.~4\textbf{D}, the monotonic increase of the Larmor frequency with applied bias is shown. These results indicate that electrical fields couple strongly to the spin coherence, and that $T_{2}^{*}$ is actually limited by electric field fluctuations rather than nuclear hyperfine effects. We investigated possible sample-dependent effects by repeating the same experiments on different QDs, with similar results. In addition, similar values for $T_{2}^{*}$ were obtained for the QDs in the $\delta$-doped sample~\cite{SOM}.

Our measurement of $T_{2}^{*}$ contrasts markedly with an estimate of $T_{2}^{*}$ obtained via coherent population trapping in similar HH-doped QDs~\cite{holecpt}. However, that experiment effectively filters a hole precession process for one particular spectral location, removing the potential dephasing effects of spectral diffusion. Consequently, substantially longer values of $T_{2}^{*}$ are reported, possibly corresponding to the onset of nuclear-induced dephasing.

While the ability of electric fields to shift the effective QD Larmor frequency may unfortunately impact $T_{2}^{*}$, it simultaneously provides a convenient
advantage of hole pseudo-spin qubits over electron-spin qubits: namely, it introduces the ability to locally tune multiple qubits into resonance, which may have important implications for viable two-qubit gates, and aid in scalability to many-qubit systems.

Finally, we use a spin-echo technique~\cite{daveT2} for measuring the $T_{2}$-decoherence time of the HH qubit. Fig.~4\textbf{E} illustrates the fringes obtained from scanning the $\pi$-echo-pulse, and Fig.~4\textbf{F} shows the fringe contrast as a function of total delay 2$T$. The decay is best fit by an exponential (i.e., $\propto\exp[-(t/T_2)]$), with $T_{2}=1.1$~$\mu$s, the same order of magnitude as for electron spins~\cite{daveT2}. Here as well, magnetic-field-dependent studies do not show any dependence on the field between 6 and 10~T. In combination with a single-qubit operation time of about 20 ps, this $T_2$-value allows for approximately 50,000 operations within the coherence time of the qubit. Although 500 times longer than $T_{2}^{*}$, the obtained $T_{2}$-value is still lower than what is theoretically expected for a nuclear-spin-limited decay. Phonon interactions are expected to only weakly affect the quantum dot hole spin~\cite{Bulaev,fischerprb,holeT1}, and can therefore also be excluded as a dominant source of decoherence. In addition, $T_{1}$ in our experiment, while limited by leakage of the optical pump laser, was measured to be at least 1 to 2 orders of magnitude larger~\cite{SOM}. It therefore appears that $T_{2}$ is limited by a similar, non-nuclear mechanism as that which most likely limits $T_{2}^{*}$, i.e. charge-induced spectral diffusion. Such a process can be mimicked by AC-modulation of the external voltage bias, and we have indeed observed a suppression of $T_{2}$ when introducing such a modulation. We measured significant dot-to-dot variance of $T_{2}$, sometimes measuring $T_{2}$ as low as several hundred ns. The variation is likely due to differences in the spin-orbit contribution to the hole-pseudospin Hamiltonian, in large part due to different HH-LH mixing for dots of different shapes and levels of strain. Further understanding the decoherence mechanisms for holes may enable extension of the spin coherence through further device engineering, as well as through the application of advanced dynamical decoupling schemes, as have recently been demonstrated for electron spins~\cite{bluhmdd}.

In conclusion, we experimentally confirm the recent theoretical predictions~\cite{Bulaev,fischerprb, Fischermixed} of the weak interactions of III-V QD hole spins with the nuclear bath. Our experiments also show that ultrafast coherent control techniques work with high fidelity for hole qubits, resulting in coherence times comparable to those for electron spins, and that complex nuclear spin bath dynamics no longer measurably affect the qubit. However, further study of non-nuclear decoherence mechanisms is required to fully exploit the promise of QD holes as qubits.

\nocite{acks}

\bibliographystyle{apsrev}
\bibliography{Holebib}

\begin{figure}\begin{center}
\includegraphics[width=\textwidth]{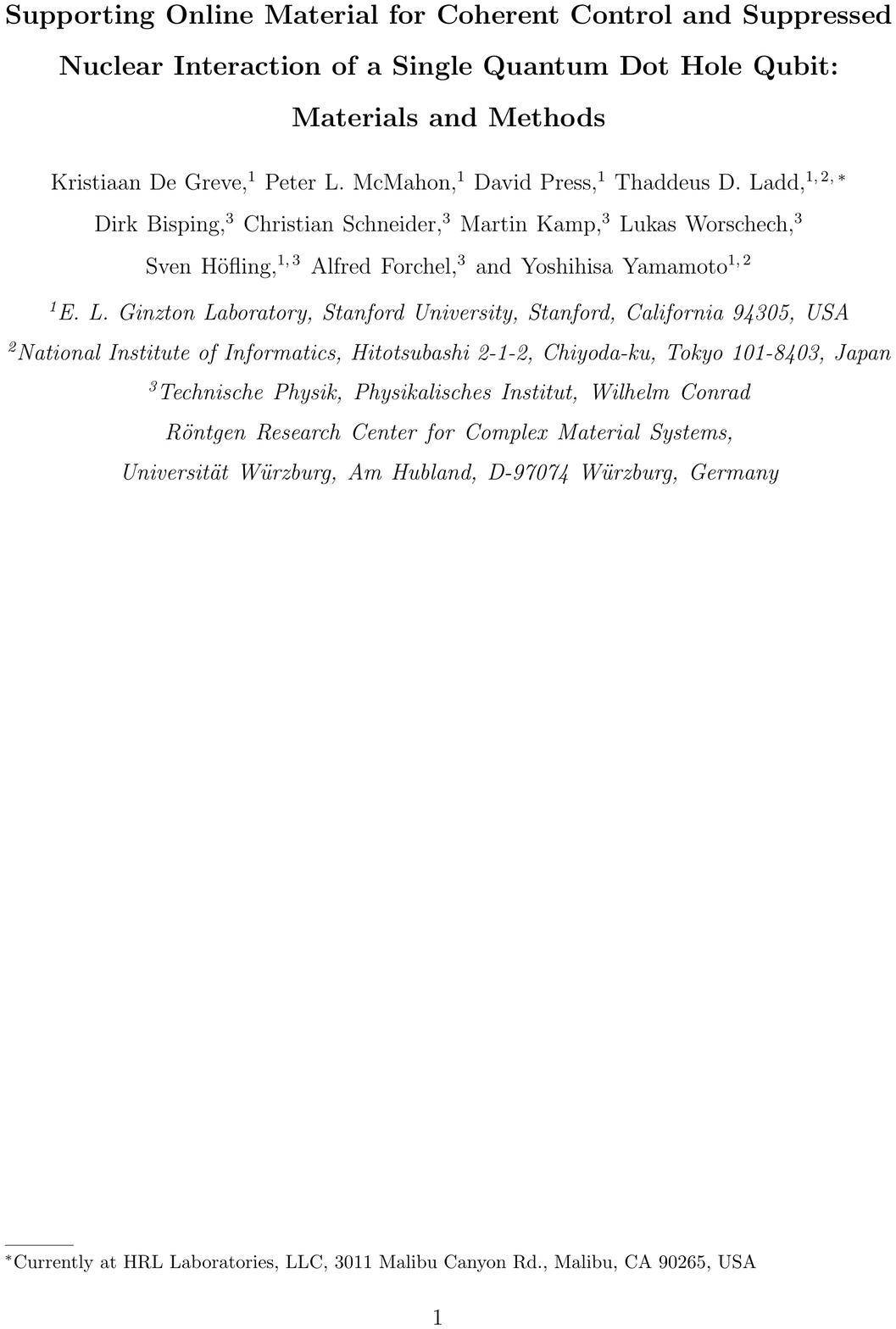}
\end{center}
\end{figure}

\begin{figure}\begin{center}
\includegraphics[width=\textwidth]{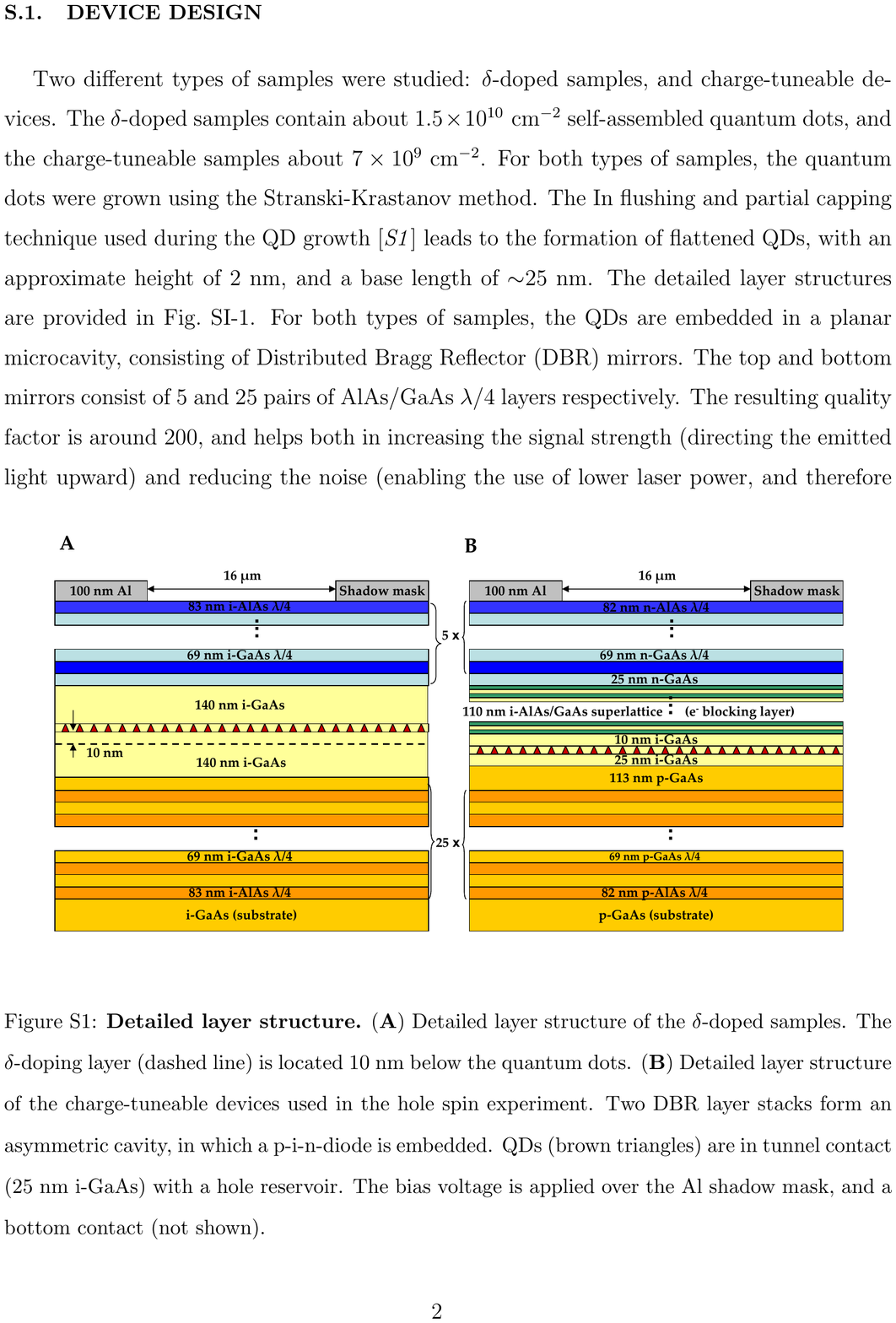}
\end{center}
\end{figure}

\begin{figure}\begin{center}
\includegraphics[width=\textwidth]{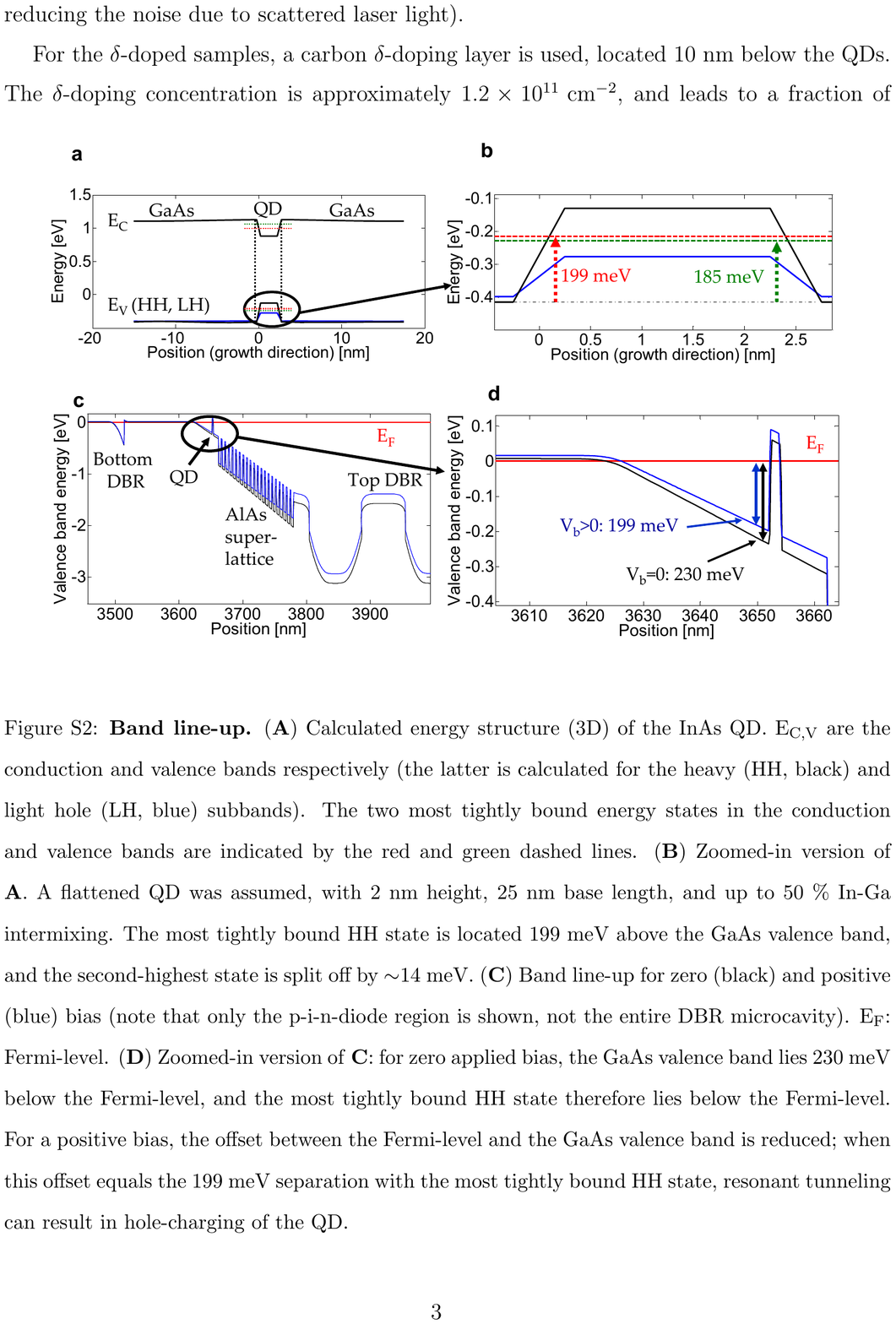}
\end{center}
\end{figure}

\begin{figure}\begin{center}
\includegraphics[width=\textwidth]{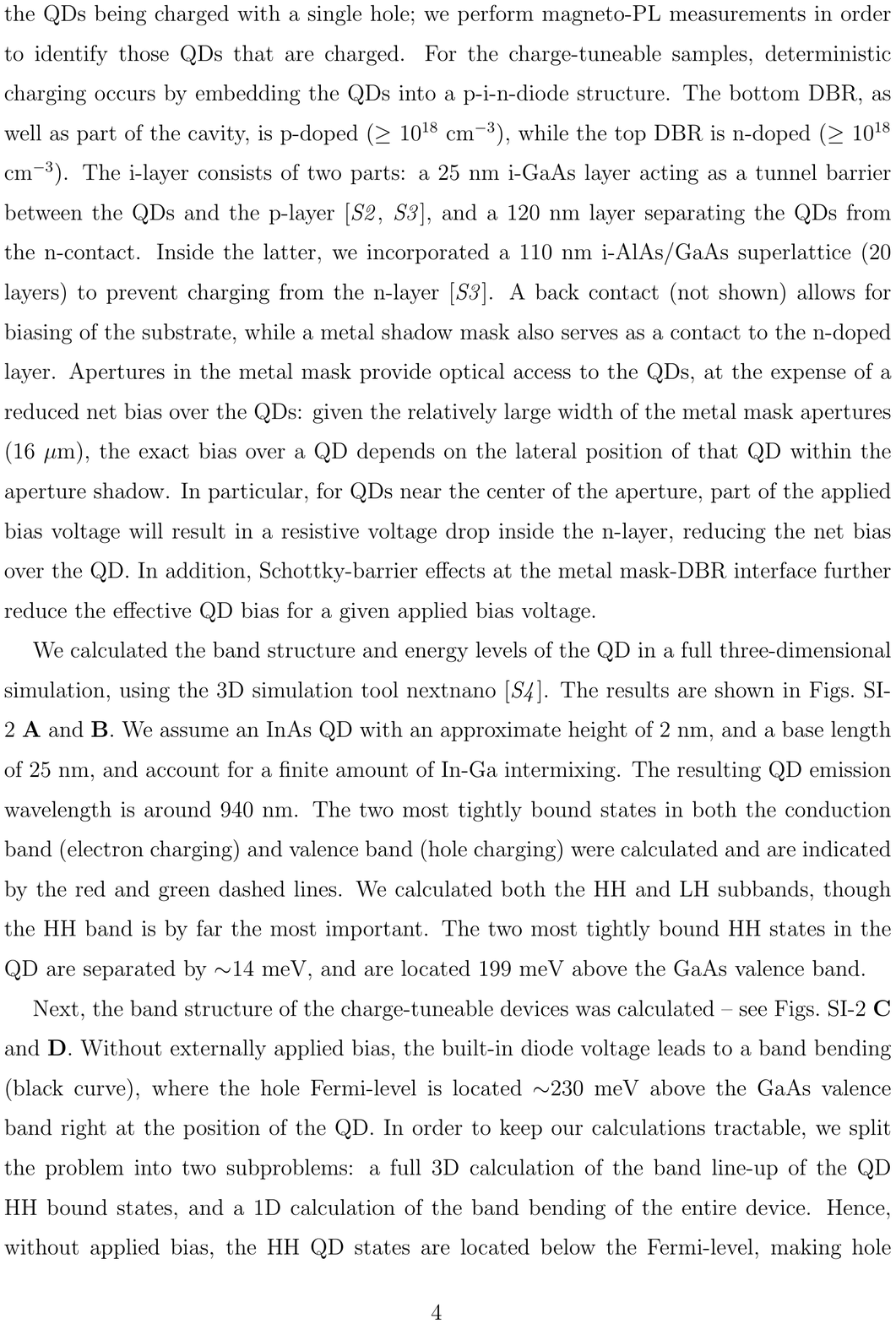}
\end{center}
\end{figure}

\begin{figure}\begin{center}
\includegraphics[width=\textwidth]{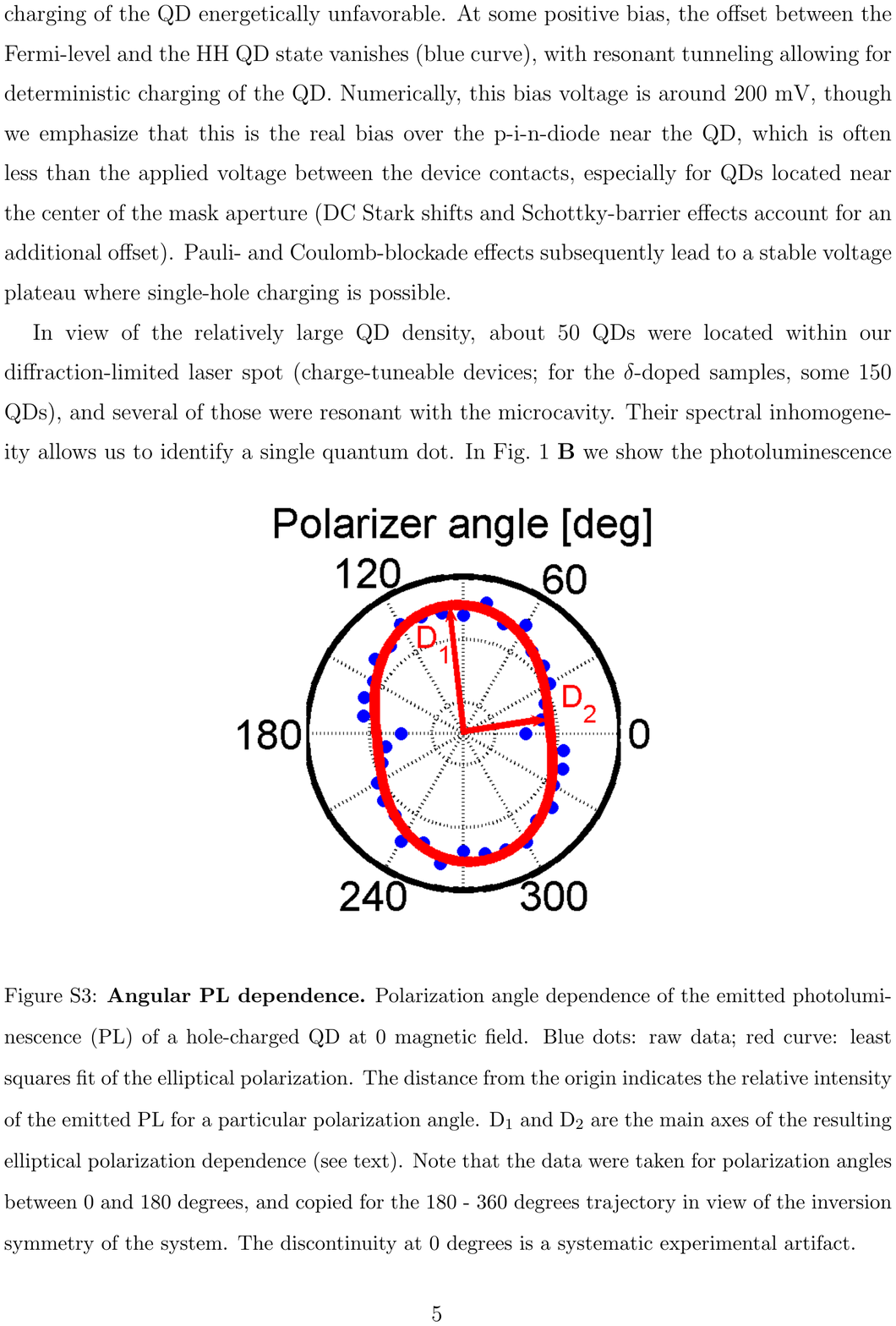}
\end{center}
\end{figure}

\begin{figure}\begin{center}
\includegraphics[width=\textwidth]{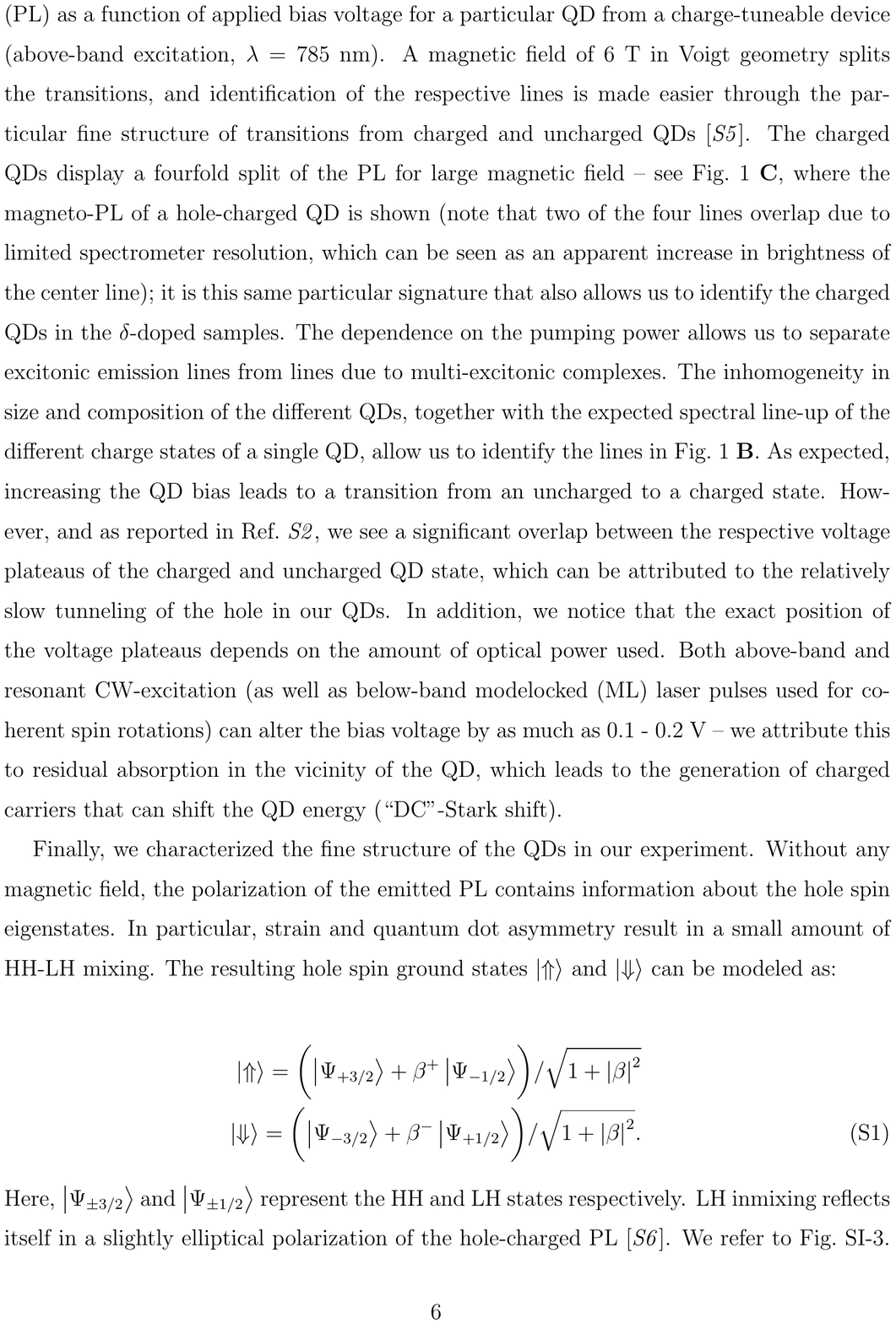}
\end{center}
\end{figure}

\begin{figure}\begin{center}
\includegraphics[width=\textwidth]{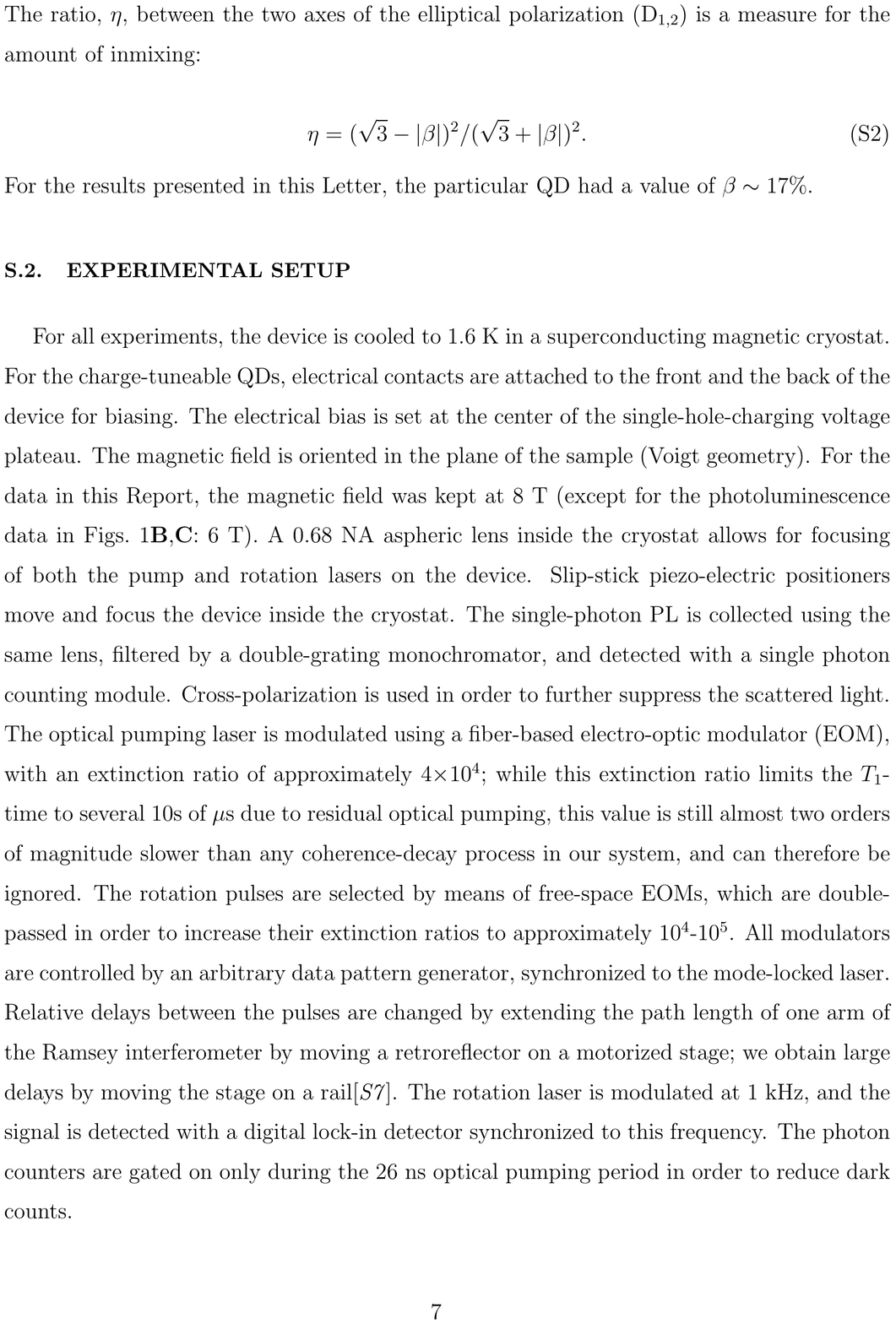}
\end{center}
\end{figure}

\begin{figure}\begin{center}
\includegraphics[width=\textwidth]{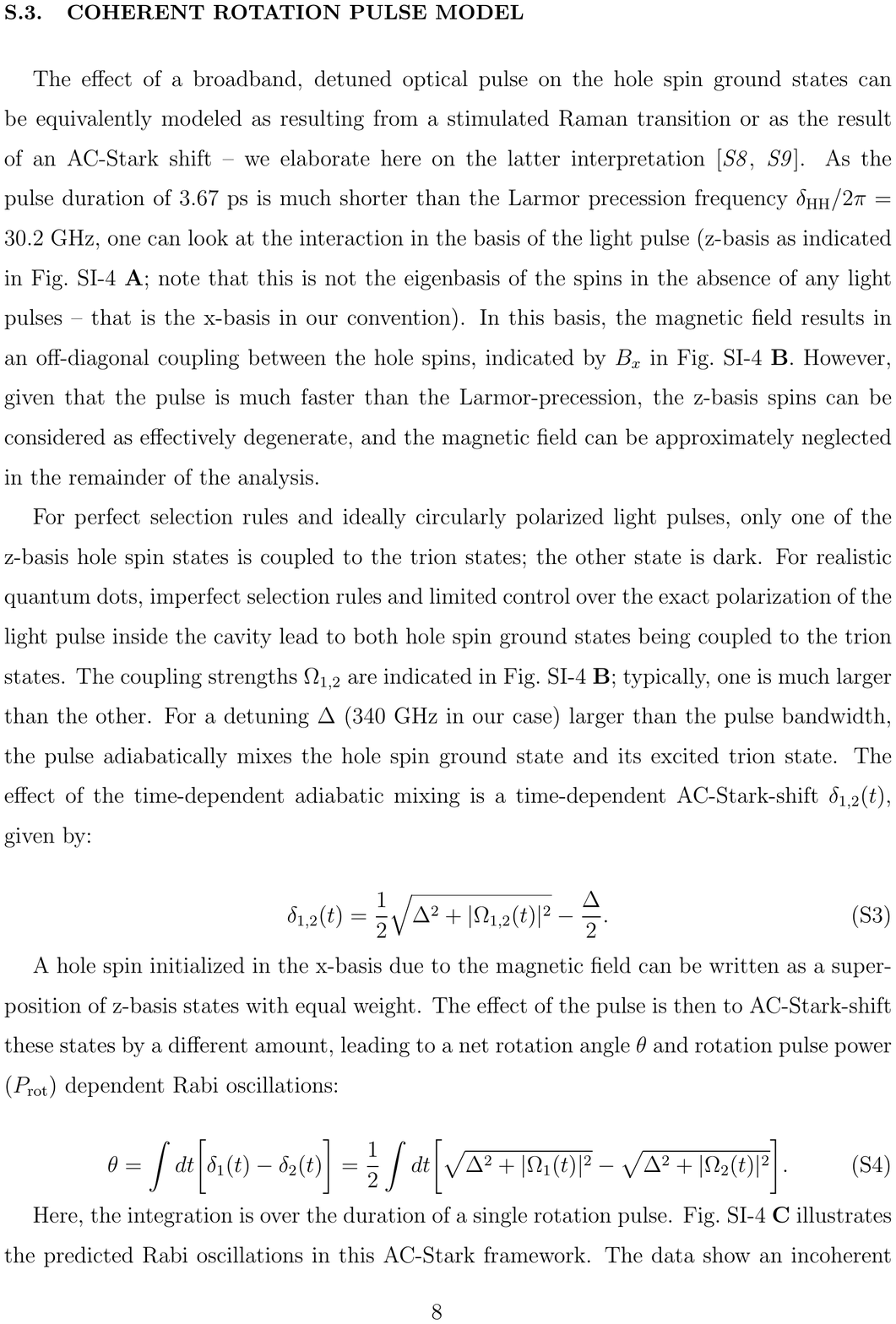}
\end{center}
\end{figure}

\begin{figure}\begin{center}
\includegraphics[width=\textwidth]{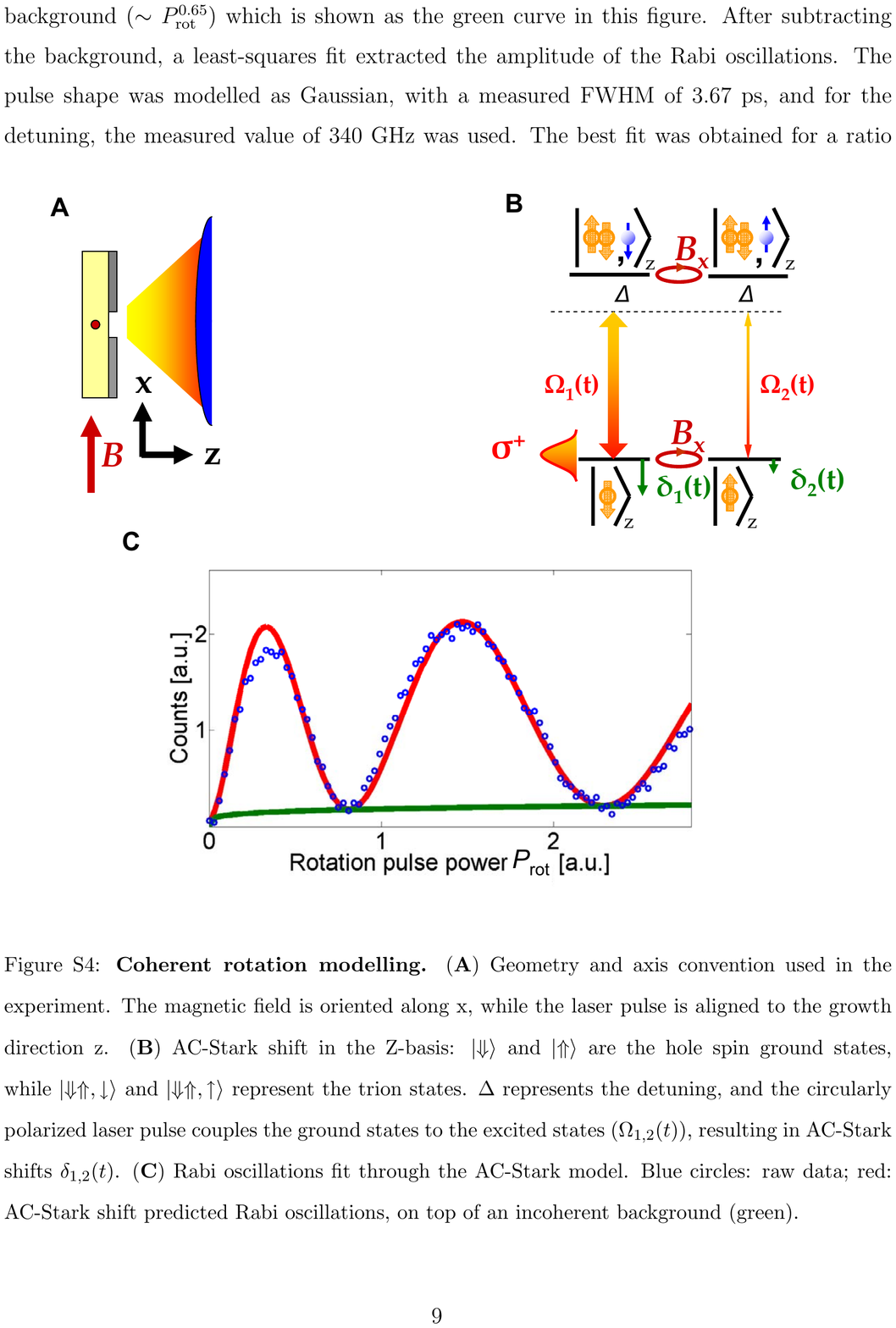}
\end{center}
\end{figure}

\begin{figure}\begin{center}
\includegraphics[width=\textwidth]{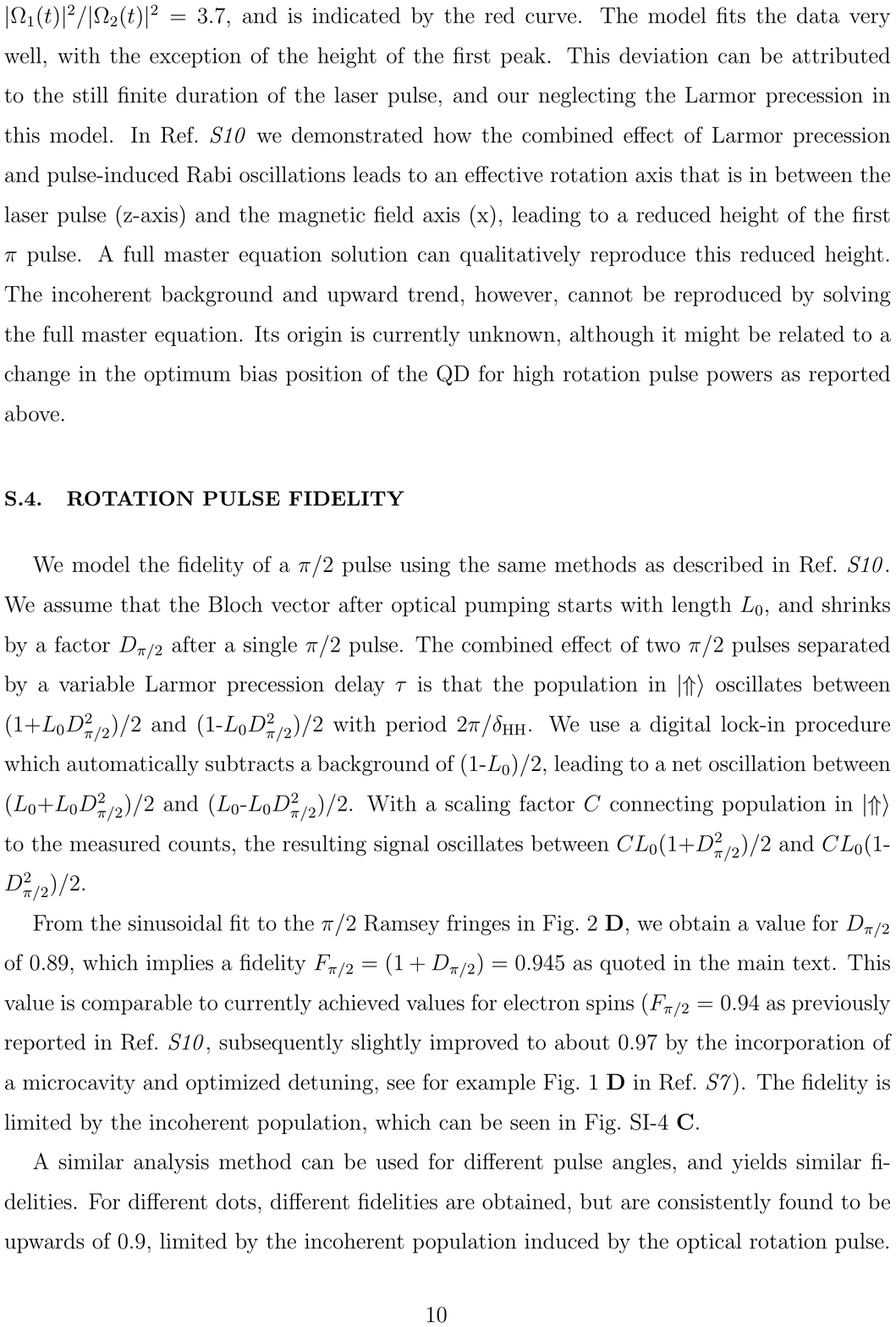}
\end{center}
\end{figure}

\begin{figure}\begin{center}
\includegraphics[width=\textwidth]{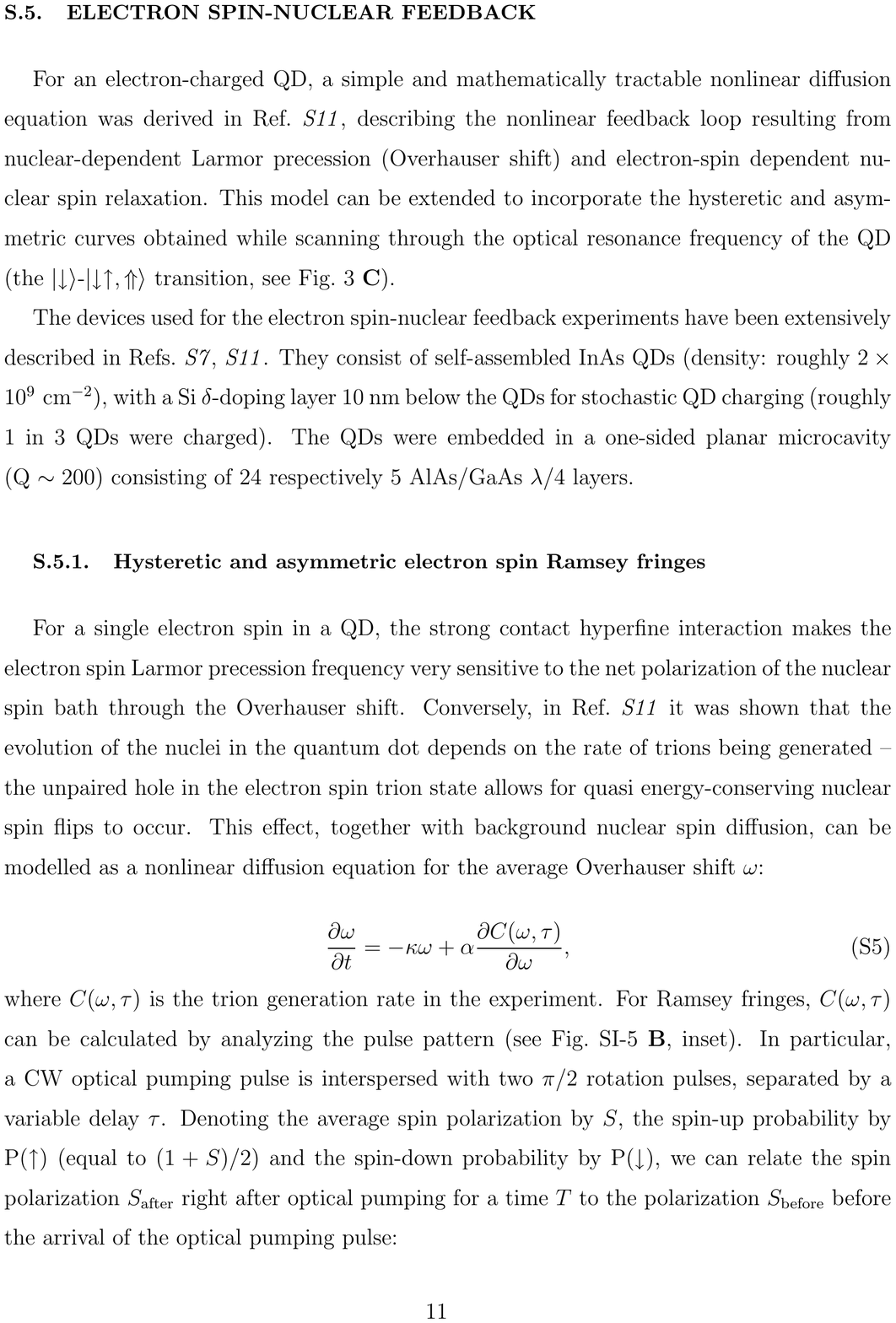}
\end{center}
\end{figure}

\begin{figure}\begin{center}
\includegraphics[width=\textwidth]{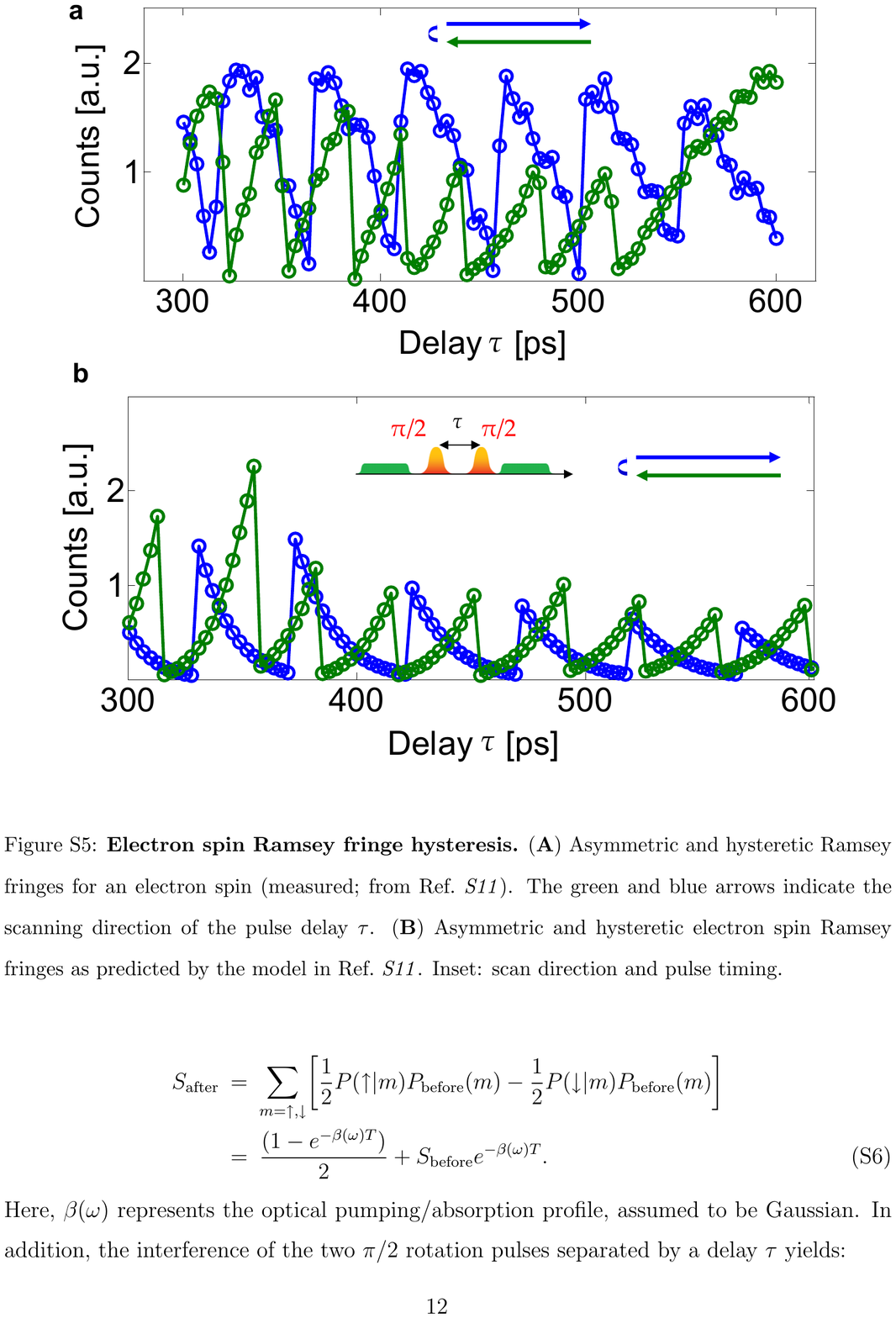}
\end{center}
\end{figure}

\begin{figure}\begin{center}
\includegraphics[width=\textwidth]{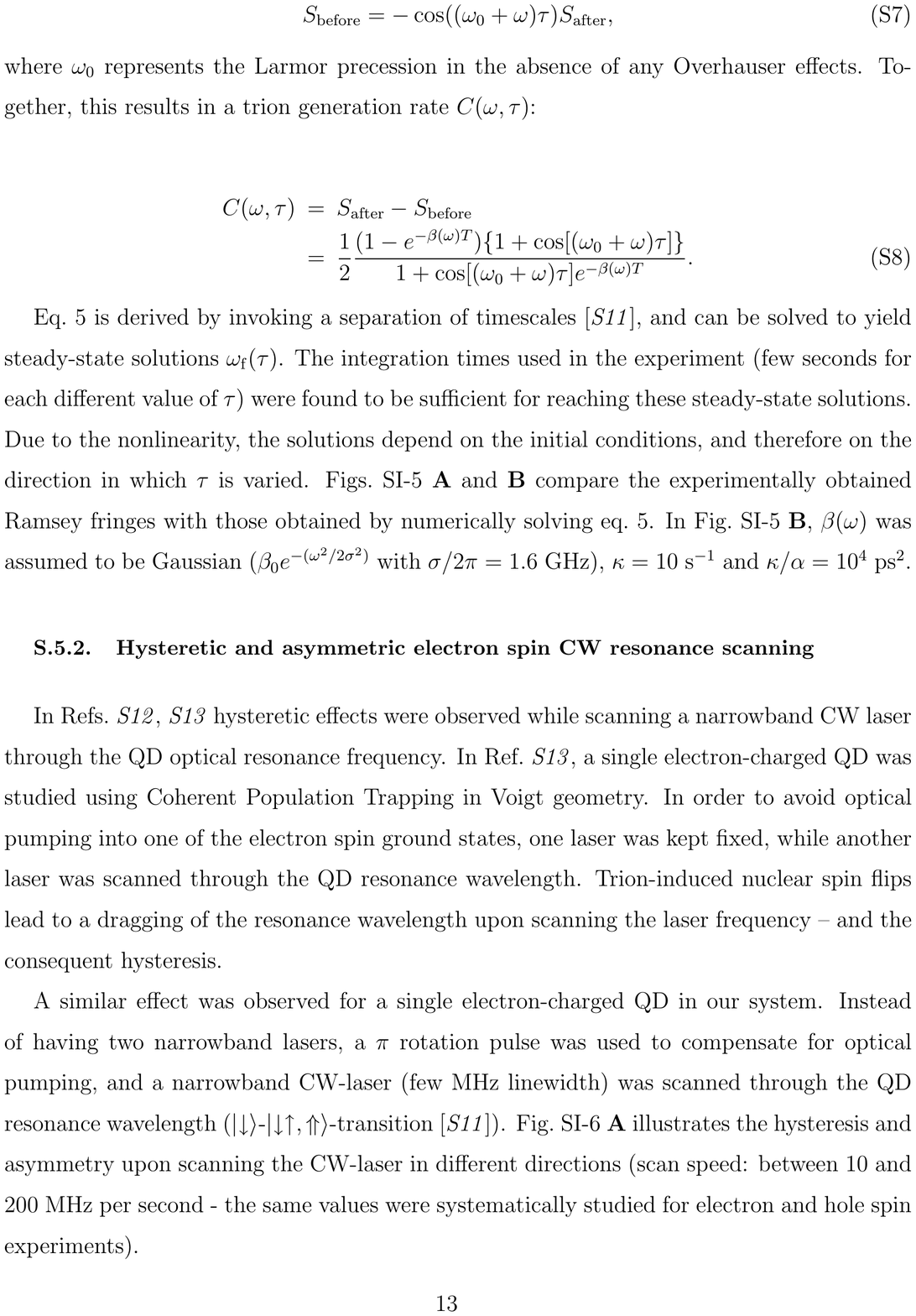}
\end{center}
\end{figure}

\begin{figure}\begin{center}
\includegraphics[width=\textwidth]{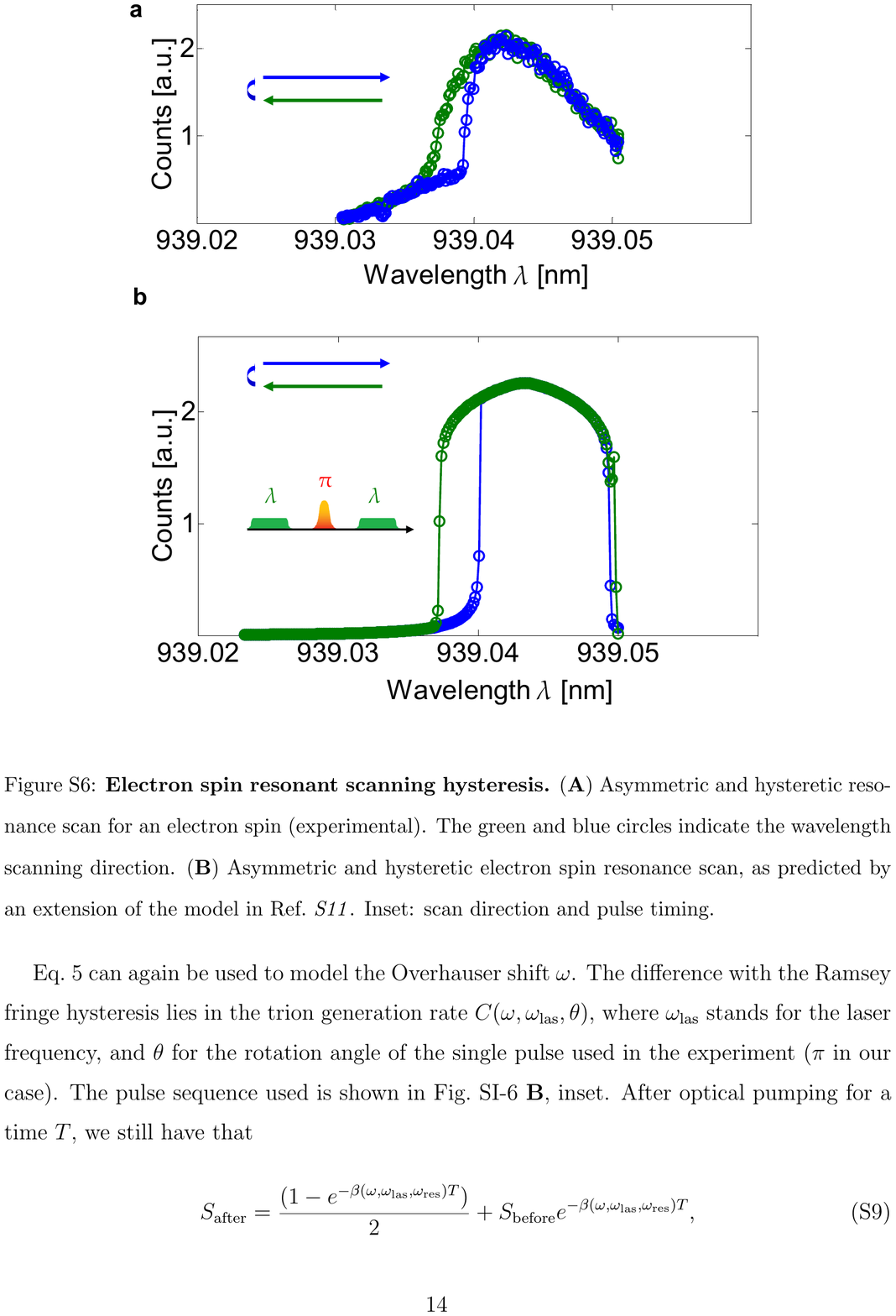}
\end{center}
\end{figure}

\begin{figure}\begin{center}
\includegraphics[width=\textwidth]{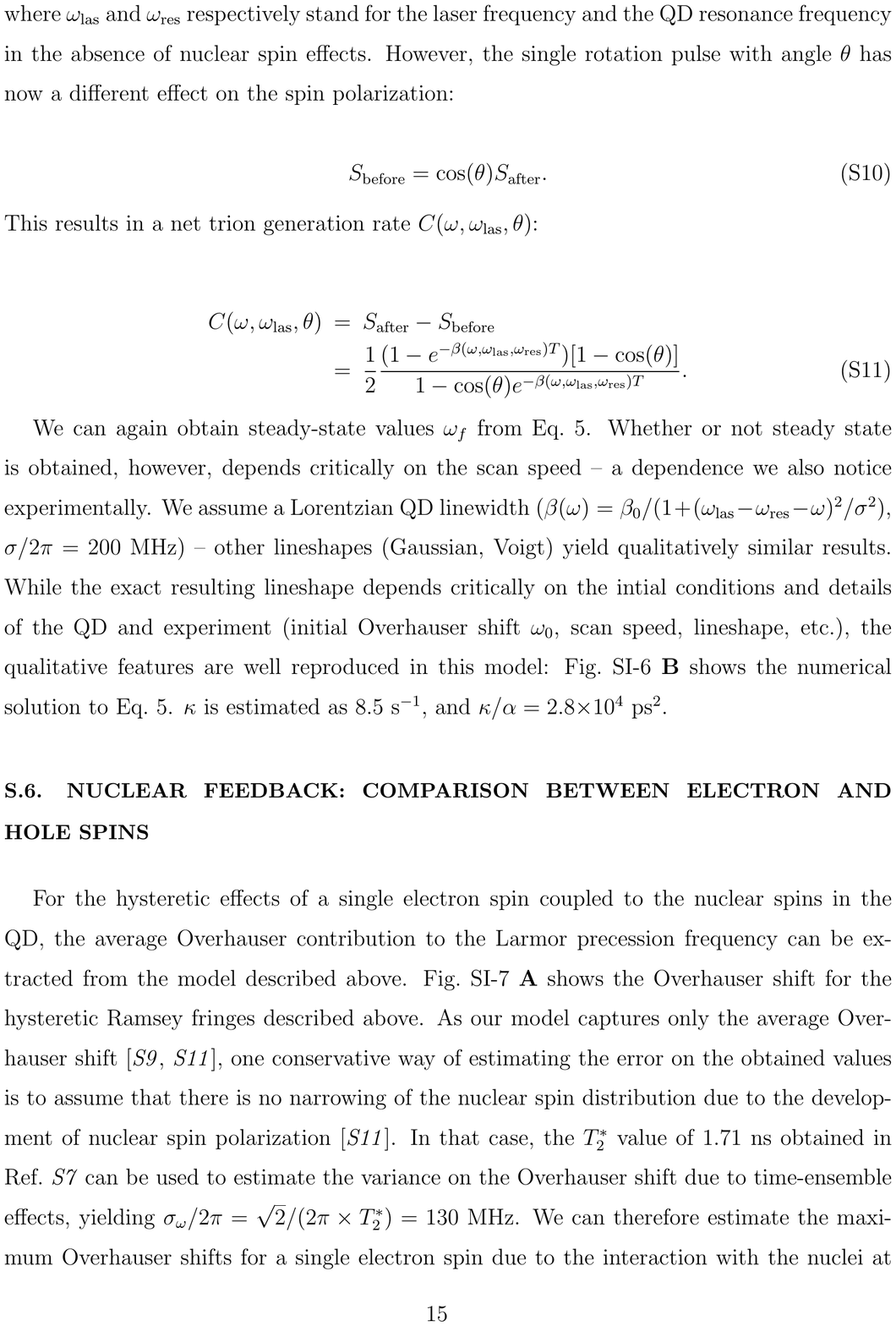}
\end{center}
\end{figure}

\begin{figure}\begin{center}
\includegraphics[width=\textwidth]{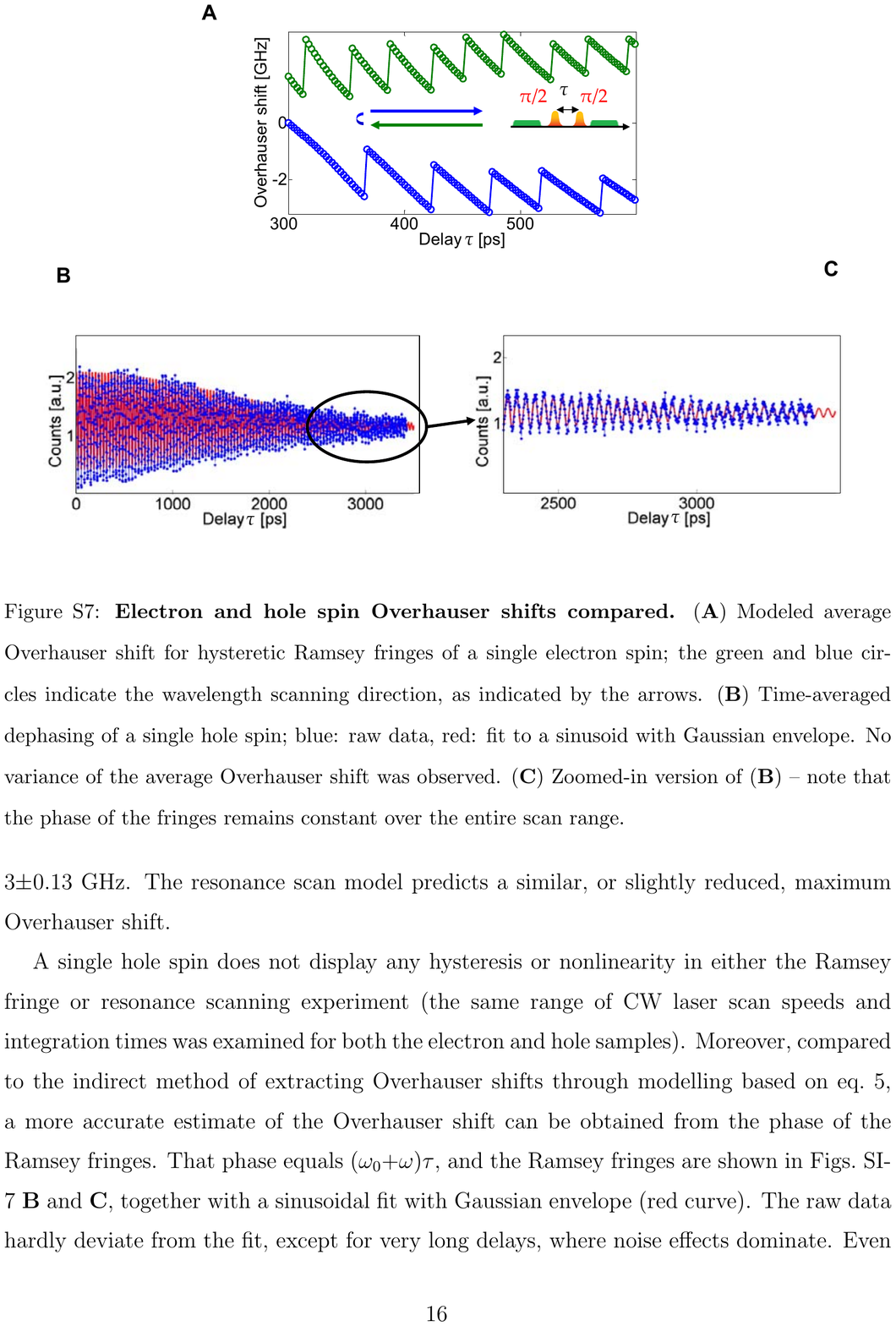}
\end{center}
\end{figure}

\begin{figure}\begin{center}
\includegraphics[width=\textwidth]{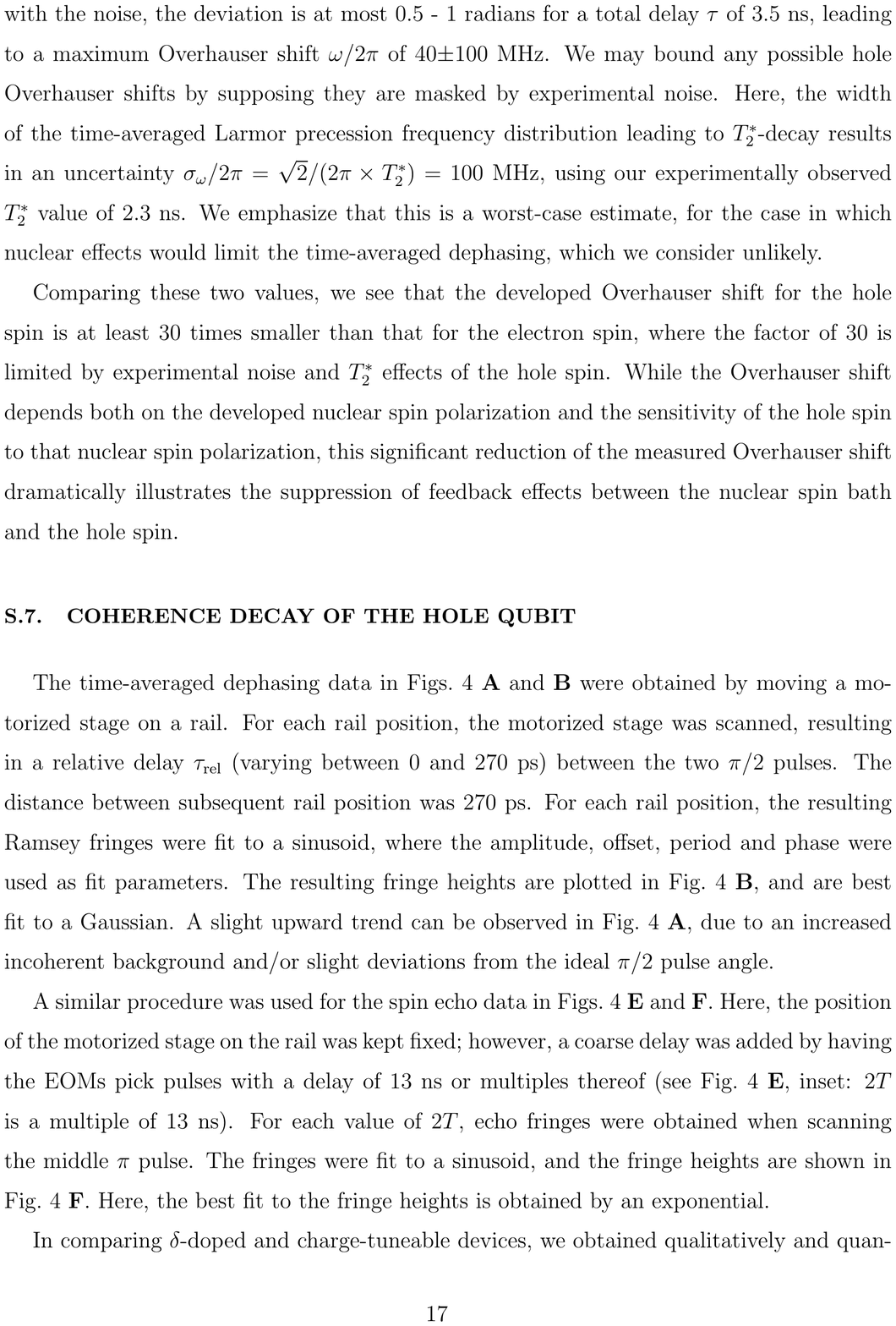}
\end{center}
\end{figure}

\begin{figure}\begin{center}
\includegraphics[width=\textwidth]{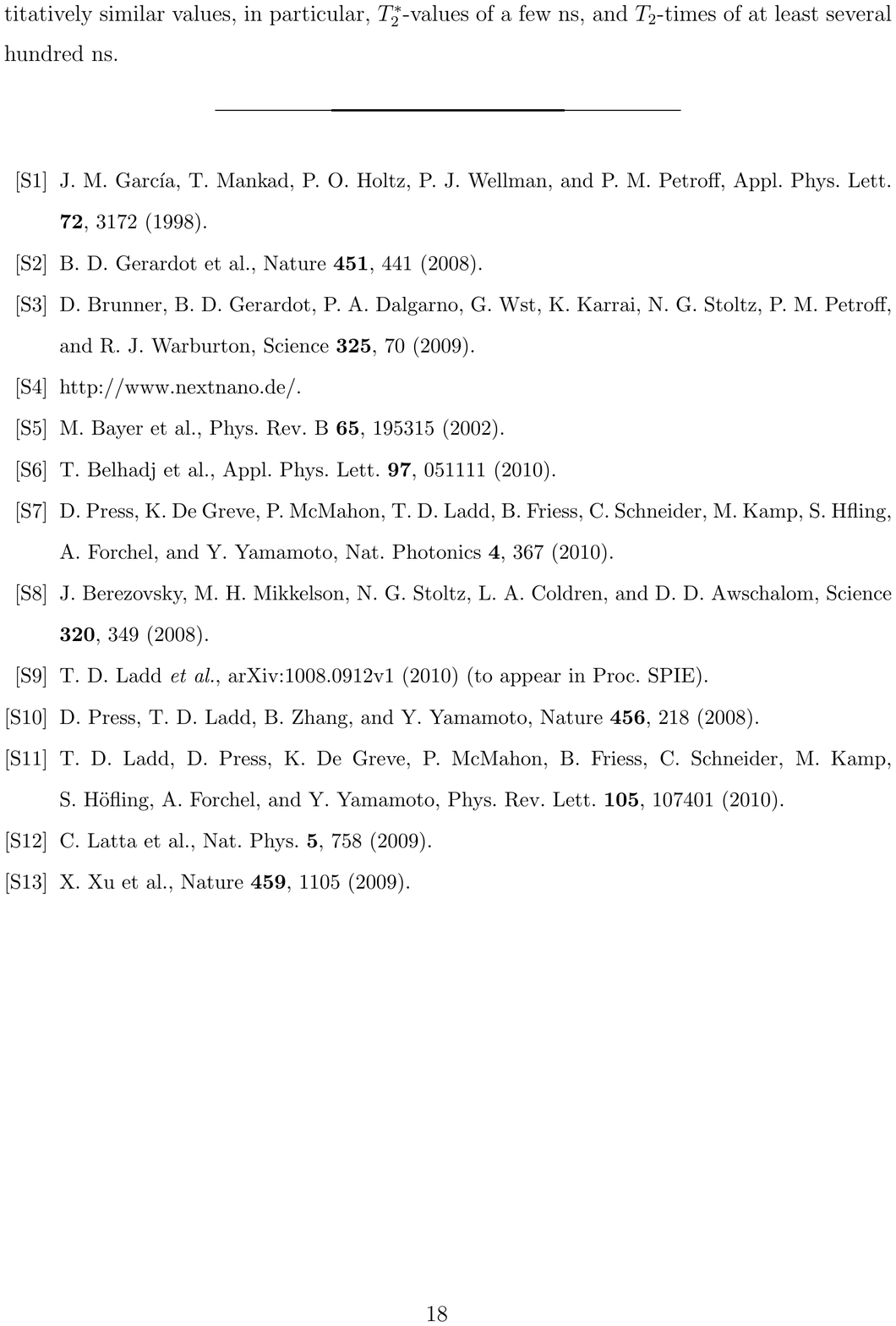}
\end{center}
\end{figure}
\end{document}